\documentclass{pasj00}
\draft

\begin{document}
\SetRunningHead{H. Asada}{Formula to Determine All Binary Elements} 
\Received{}
\Accepted{}

\title{Note on Inversion Formula 
to Determine Binary Elements by Astrometry} 

\author{Hideki \textsc{Asada}} %
\affil{Faculty of Science and Technology, Hirosaki University, 
Hirosaki, Aomori 036-8561}
\email{asada@phys.hirosaki-u.ac.jp}


%


\maketitle

\begin{abstract}
Simplified solutions to determine binary elements by astrometry 
were obtained in terms of elementary functions 
(Asada et al. 2004), 
and therefore require neither iterative nor numerical methods. 
In the framework of the simplified solution, 
this paper discusses the remaining two parameters of 
the time of periastron passage 
and the longitude of ascending node 
in order to complete the solution. 
We thus clarify a difference between the simplified solution 
and other analytical methods. 
\end{abstract}

\section{Introduction}
Recently, we developed a formulation for 
determining binary elements with astrometric observations. 
The simplified solution is written in terms of elementary functions, 
and therefore requires neither iterative nor numerical methods 
(Asada et al. 2004). 
This solution has been generalized to a binary system 
in open (hyperbolic or parabolic) orbits 
as well as closed (elliptic) ones (Asada 2007). 
An extension to observational data has been also discussed 
(Asada et al. 2007). 
The solution gives an explicit form of binary elements 
such as the eccentric anomaly and the major axis of 
elliptic orbits. 
Oyama et al. (2008) made an attempt to use this solution 
for discussing some uncertainty in binary elements 
because of large scatter of their data points, 
when they measured proper motions of maser sources 
in the galactic center with VERA. 

On the other hand, the remaining parameters of 
the time of periastron passage and the longitude of ascending node 
are not discussed in the simplified solution.  
Hence, the solution is rather simplified. 
However, these parameters are needed to make a comparison 
between the simplified solution and conventional ones. 
In addition, the lack of information on the remaining parameters 
apparently suggests a certain incompleteness of the simplified solution. 
In this brief article, therefore, we shall derive, 
in the framework of the simplified solution, 
both the time of periastron passage and 
the longitude of ascending node 
in order to complete the solution. 

Astrometry plays a fundamental role in astronomy 
through providing useful star catalogs based on 
precise measurements of the positions and movements of stars and 
other celestial bodies. 
For instance, 
astrometric observations provide an useful method of 
determining mass of various unseen celestial objects 
currently such as a massive black hole (Miyoshi et al. 1995),  
an extra-solar planet (Benedict et al. 2002) 
and two new satellites of Pluto (Weaver et al. 2006). 
Astrometry of Sharpless 269 with VERA detects a trigonometric 
parallax corresponding to a distance of 5.28 kpc, 
which is the smallest parallax ever measured, 
and puts the strongest constraint on 
the flatness of outer rotation curve (Honma et al. 2007). 
Accordingly, astrometry has attracted renewed interests, 
since the Hipparcos mission successfully provided  
us the precise catalog at the level of a milliarcsec. 
In fact, 
there exist several projects of space-borne astrometry 
aiming at a accuracy of a few microarcseconds,  
such as SIM\footnote{http://sim.jpl.nasa.gov/} (Shao 2004), 
GAIA\footnote{hthttp://www.rssd.esa.int/index.php?project=GAIA\&page=index} 
(Mignard 2004, Perryman 2004) and 
JASMINE\footnote{http://www.jasmine-galaxy.org/} 
(Gouda et al. 2007).

In this paper, we focus on an astrometric binary, 
for which only one of the component stars can be visually observed 
but the other cannot, like a black hole or a very dim star. 
In this case, it is impossible 
to directly measure the relative vector connecting the two objects, 
because the secondary is not directly observed. 
The position of the star is repeatedly measured relative to 
reference stars or quasars. 
On the other hand, 
the orbit determination of resolved double stars (visual binaries), 
which are a system of two visible stars,  
was solved first by Savary in 1827 
and by many authors including Kowalsky, Thiele and Innes 
(Binnendijk 1960, Aitken 1964 for a review on earlier works; for the
state-of-the-art techniques, e.g., Eichhorn and Xu 1990, 
Catovic and Olevic 1992, Olevic and Cvetkovic 2004). 
The relative vector from the primary star to the secondary has 
an elliptic motion with a focus at the primary. 
This relative vector is observable only for resolved double stars.

In conventional methods of orbit determination, 
the time of periastron passage is one of important parameters 
because it enters the Kepler's equation as 
\begin{equation}
t = t_0 + \frac{T}{2\pi}(E-e_K\sin{E}), 
\label{Kepler}
\end{equation}
where $t_0$, $T$, $e_K$ and $E$ denote 
the time of periastron passage, orbital period, eccentricity and 
eccentric anomaly, respectively 
(e.g., Danby 1988, Roy 1988, Murray and Dermott 1999, 
Beutler 2004). 
The simplified solution does not use the Kepler's equation 
in order to avoid treating such a transcendental equation. 

This paper is organized as follows. 
Our notation in the simplified solution will be summarized in $\S$ 2. 
The time of periastron passage in the simplified solution 
will be derived in $\S$ 3. 
The longitude of ascending node will be obtained in $\S$ 4.

\section{Simplified solution: Our notation} 
Our notation in the simplified solution 
is briefly summarized as follows. 
We neglect motions of the observer and the common center 
in our galaxy. 
Namely, we take account only of the Keplerian motion of a star around 
the common center of mass of a binary system. 
Let us define $(x, y)$ as the Cartesian coordinates 
on a celestial sphere, in such a way that 
the apparent (observed) ellipse on the celestial sphere 
can be expressed in the standard form as 
\begin{equation}
\frac{x^2}{a^2}+\frac{y^2}{b^2}=1 , 
\label{ellipse3}
\end{equation}
where $a\geq b$. 
The eccentricity $e$ is $\sqrt{1-b^2/a^2}$. 
This eccentricity may be different from $e_K$, 
the eccentricity of the actual elliptic orbit, 
because of the inclination of the orbital plane 
with respect to the line of our sight. 
The star is located at $P_j=(x_j, y_j)$ 
on the celestial sphere 
at the time of $t_j$ for $j=1, \cdots, n$. 

\begin{figure}
  \begin{center}
    \FigureFile(120mm,120mm){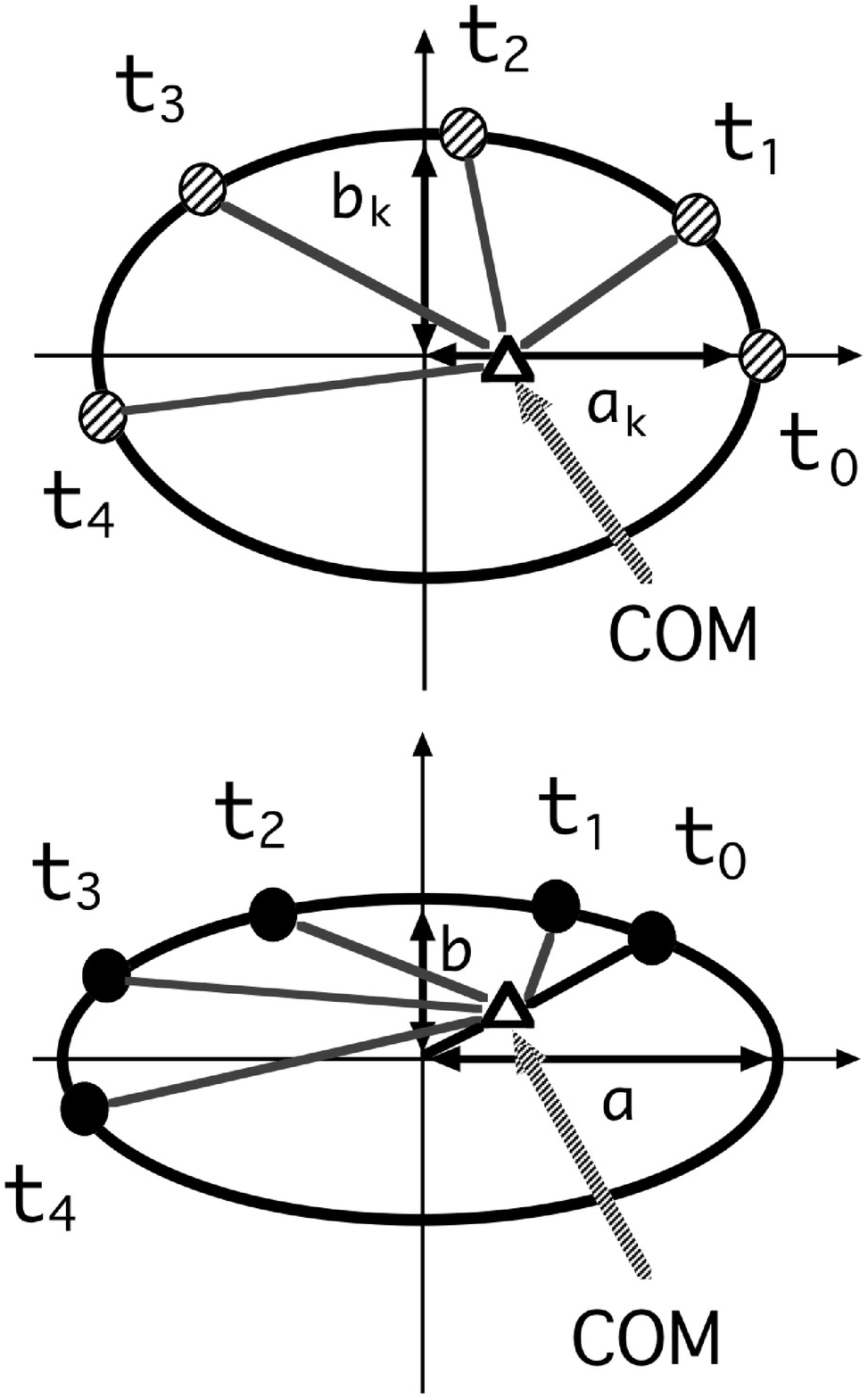}
  \end{center}
  \caption{Schematic figures of 
the actual elliptical orbit (top figure)
and the apparent one (bottom one). 
The semimajor and semiminor axes of the actual elliptical orbit 
are denoted by $a_k$ and $b_k$, respectively. 
Those of the apparent one are $a$ and $b$, respectively.
The position of the star at each time is denoted 
by shaded circles (in the top figure) and 
filled circles (in the bottom figure). 
The triangles indicate the center of mass (COM). 
In the bottom figure, the projected center of mass 
is located on the line connecting the center of the apparent ellipse 
and the projected periastron (at $t_0$).}
\label{fig1}
\end{figure}

We use a fact that the law of constant-areal velocity 
still holds, even after a Keplerian orbit is projected 
onto the celestial sphere.  
Here, the area is swept by the line interval 
between the star and the projected common center of mass 
but not a focus of the apparent ellipse 
(See Fig. $\ref{fig1}$). 
This fact is expressed as 
\begin{equation}
\frac{S}{T}=\frac{S(k, j)}{T(k, j)} ,
\label{areavelocity}
\end{equation}
where $S(k, j)$ and $S$ denote the area swept during 
the time interval, $T(k, j)=t_k-t_j$ for $t_k > t_j$, 
and the total area of the apparent
ellipse $\pi a b$, respectively.
The swept area is expressed as (Asada et al. 2004, Asada 2007) 
\begin{equation}
S(k, j)=\frac12 ab  
\Bigl[
u_k-u_j
-\frac{x_e}{a}(\sin u_k-\sin u_j)
+\frac{y_e}{b}(\cos u_k-\cos u_j) 
\Bigr] . 
\label{areaS}
\end{equation}
The eccentric anomaly in the apparent ellipse is given 
by $u_j=\arctan(ay_j/bx_j)$.

The orbital elements can be expressed explicitly as elementary
functions of the locations of four observed points and 
their time intervals (Asada et al. 2004).   
Let us take four observed points 
$P_1$, $P_2$, $P_3$ and $P_4$ for $t_1<t_2<t_3<t_4$. 
The location $(x_e, y_e)$ of the projected common center 
is given by
\begin{eqnarray}
x_e&=&-a \frac{F_1 G_2-G_1 F_2}{E_1 F_2-F_1 E_2} , 
\label{xe}\\
y_e&=&b \frac{G_1 E_2-E_1 G_2}{E_1 F_2-F_1 E_2} , 
\label{ye}
\end{eqnarray}
where $E_j$, $F_j$ and $G_j$ are elementary functions 
of $T(j+2, j+1)$, $T(j+1, j)$ and $u_{k}$ for $k=j$, $j+1$, $j+2$.  
The eccentric anomaly in the actual ellipse (on the orbital plane) 
is denoted as $E$ (See Eq. $(\ref{Kepler})$). 

Given $a$, $b$, $x_e$ and $y_e$, 
we can analytically determine 
the parameters $e_K$, $i$, $a_K$ and $\omega$ 
 as (Asada et al. 2004) 
\begin{eqnarray}
&&e_K=\sqrt{\frac{x_e^2}{a^2}+\frac{y_e^2}{b^2}} ,  
\label{eK}\\
&&\cos i=\frac12 (\xi - \sqrt{\xi^2-4}) , 
\label{cosi}\\
&&a_K=\sqrt{\frac{C^2+D^2}{1+\cos^2 i}} , 
\label{aK}\\
&&\cos 2\omega=\frac{C^2-D^2}{a_K^2 \sin^2 i} , 
\label{cos2omega}
\end{eqnarray}
where 
\begin{eqnarray}
&&C=\frac{1}{e_K}\sqrt{x_e^2 + y_e^2} , 
\label{C2}\\
&&D=\frac{1}{abe_K}\sqrt{\frac{a^4 y_e^2 + b^4 x_e^2}{1-e_K^2}} , 
\label{D2}\\
&&\xi=\frac{(C^2+D^2)\sqrt{1-e_K^2}}{ab} . 
\label{xi}
\end{eqnarray}

\section{Time of periastron passage}
In order to determine $a_K$ and $e_k$ 
for an actual ellipse, 
the simplified solution requires neither the time of periastron passage 
$t_0$ nor the longitude of ascending node, $\Omega$ (Asada et al. 2004). 
If one wishes to know $t_0$ and $\Omega$, however, they can be determined 
as follows (See also Fig. $\ref{fig2})$. 
First, we discuss $t_0$ in this section. 

The projected position of the periastron on the celestial sphere,
${\bf P}_A$, is determined as 
\begin{equation}
{\bf P}_A = \frac{1}{e_K} (x_e, y_e) , 
\label{PA-1}
\end{equation}
because the ratio of the semimajor axis to the distance 
between the center and the focus of the ellipse remains unchanged, 
even after the projection (Asada et al. 2004).
The eccentric anomaly $u_A$ of the periastron 
in the apparent ellipse is introduced as 
\begin{equation}
{\bf P}_A=(a \cos u_A, b \sin u_A) , 
\label{PA-2}
\end{equation}
where ${\bf P}_A$ is given also by Eq. ($\ref{PA-1}$). 
Thereby, we can determine $u_A$ (mod $2\pi$). 

By using Eq. ($\ref{areavelocity}$), we obtain 
\begin{equation}
\frac{S(1, 0)}{T(1, 0)} = \frac{S(2, 1)}{T(2, 1)} , 
\label{area}
\end{equation}
where we can determine $S(1, 0)$ because the eccentric anomaly 
in the apparent ellipse at $t_0$,  
denoted as $u_0$, is nothing but $u_A$, 
which has been determined by Eqs. ($\ref{PA-1}$) and ($\ref{PA-2}$). 
Therefore, Eq. $(\ref{area})$ is solved for $t_0$ as 
\begin{equation}
t_0 = \frac{S(2, 0)}{S(2, 1)}t_1 - \frac{S(1, 0)}{S(2, 1)}t_2 . 
\label{t0}
\end{equation}
where the R.H.S. is obtained from observed quantities.

\section{Longitude of ascending node} 
Let us consider the projected periastron at ${\bf P}_A$ 
on the apparent ellipse. 
In the simplified solution, ${\bf P}_A$ is expressed 
as Eq. ($\ref{PA-1}$). 
Here we make a translation of $(x, y)$ in such a way that 
the common center of mass can be located at the origin of new coordinates 
$(x^{\prime}, y^{\prime})$. 
Namely, the $x^{\prime}$ axis is taken to lie along the major axis 
of the apparent ellipse in the celestial sphere, 
and the $y^{\prime}$ axis is perpendicular to the $x^{\prime}$ axis 
in the celestial sphere (See Fig. $\ref{fig3}$). 
In the coordinates $(x^{\prime}, y^{\prime})$, 
the position of the projected periastron becomes 
\begin{equation}
{\bf P}_A = \frac{1-e_K}{e_K} (x_e, y_e) .  
\label{PA-prime}
\end{equation}
On the other hand, by projecting the actual ellipse 
onto the celestial sphere, we obtain 
\begin{equation}
{\bf P}_A = (a_K(1-e_K) \cos\omega, a_K(1-e_K) \sin \omega \cos i) , 
\label{PA-bar}
\end{equation}
where the coordinates $(\bar{x}, \bar{y})$ are chosen 
so that the ascending node can be in the $\bar{x}$-direction 
(See Fig. $\ref{fig3}$). 

The longitude of ascending node, which is the angle between 
the $\bar{x}$ and $x^{\prime}$ axes, relates the two coordinates of 
$(x^{\prime}, y^{\prime})$ and $(\bar{x}, \bar{y})$ by rotation. 
Therefore, from Eqs. $(\ref{PA-prime})$ and $(\ref{PA-bar})$, we obtain 
\begin{eqnarray}
&&
\left(
\begin{array}{cc}
\cos\Omega & -\sin\Omega \\
\sin\Omega & \cos\Omega \\
\end{array}
\right)
\left(
\begin{array}{c}
a_K (1-e_K) \cos\omega \\
a_K (1-e_K) \sin\omega \cos i \\
\end{array}
\right)
=\frac{1-e_K}{e_K} 
\left(
\begin{array}{c}
x_e \\
y_e \\
\end{array}
\right) . 
\end{eqnarray}
This relation determines $\Omega$ (mod $2\pi$). 
For instance, we obtain explicitly 
\begin{equation}
\tan\Omega = \frac{y_e \cos\omega - x_e \sin\omega \cos i}
{y_e \sin\omega \cos i + x_e \cos\omega } , 
\label{tanOmega}
\end{equation}
where $x_e$, $y_e$, $i$ and $\omega$ in the R.H.S. 
have been determined by Eqs. ($\ref{xe}$), ($\ref{ye}$), 
($\ref{cosi}$) and ($\ref{cos2omega}$). 

In conventional methods, determining $\Omega$ is tightly coupled 
with $\omega$ and $i$. 
On the other hand, it can be done separately from 
$\omega$ and $i$ in the simplified solution. 

It should be noted that in practical applications a reference
direction chosen by observers may be different from 
the major axis of the apparent ellipse. 
In such a practical case, $\Omega$ is the angle from 
the reference direction to the direction 
of the ascending node.   
To compute the longitude of ascending node, therefore, 
the angle $\delta\Omega$ from the reference direction 
to the major axis is added into the angle measured from 
the major axis. 
In short, the longitude of ascending node is generally 
the sum of $\Omega_0$ and $\delta\Omega$, 
where $\Omega_0$ is the angle $\Omega$ determined by using 
Eq. $(\ref{tanOmega})$.   
The expression of $\delta\Omega$ is obtained 
in the straightforward manner, for instance 
as Eq. (6) in Asada et al. (2007),  
where they denoted $\delta\Omega$ as $\Omega$.

Tables 1 and 2 give an example 
to show the flow of actual determination of all six orbital elements. 
This would be helpful to the readers who 
code computing routines in practical applications 
to check their programs. 
Table 1 shows some given values for all six orbital elements 
and the orbital period. 
Based on these elements, we first prepare a virtual observation 
data set, which is listed in Table 2. 
The elements are then reproduced from the data 
by using the proposed method.  

Here, we discuss how to determine in the simplified solution 
a position of a component star 
at arbitrary time $t \equiv t_n$ (mod $T$). 
For $t_1$, $t_2$ and $t_n$, 
Eq. ($\ref{areavelocity}$) becomes 
\begin{equation}
\frac{S(n, 1)}{T(n, 1)} = \frac{S(2, 1)}{T(2, 1)} . 
\label{position}
\end{equation}
This is a transcendental equation for $u_n$ on the celestial sphere. 
This situation seems similar to that the Kepler's equation is 
transcendental in $E$ on the orbital plane.  
Here we should note that the time of periastron passage
is needed in order to treat Kepler's equation, whereas  
it is not for Eq. ($\ref{position}$). 
This is because we employ the time interval $T(k, j)$, 
while the Kepler's equation needs the time itself instead of the interval. 
Regarding this point, Thiele's method for visual binaries 
is closer to the simplified solution, in the sense that 
they use the time interval in order to delete the time of 
periastron passage. 
A crucial difference is that Thiele's method 
uses Kepler's equation on the orbital plane (Thiele 1883), 
while the simplified one does the constant areal velocity 
in the apparent ellipse on the celestial sphere.  
In this sense, the simplified solution more respects 
measured quantities on the celestial sphere than conventional ones 
(See Fig. $\ref{fig2}$). 
It is verified numerically that the above procedure 
enables us to determine in the simplified solution 
locations of a star at arbitrary time 
(See Fig. $\ref{fig4}$ for an example).

\section{Conclusion}
\begin{figure}
  \begin{center}
    \FigureFile(160mm,160mm){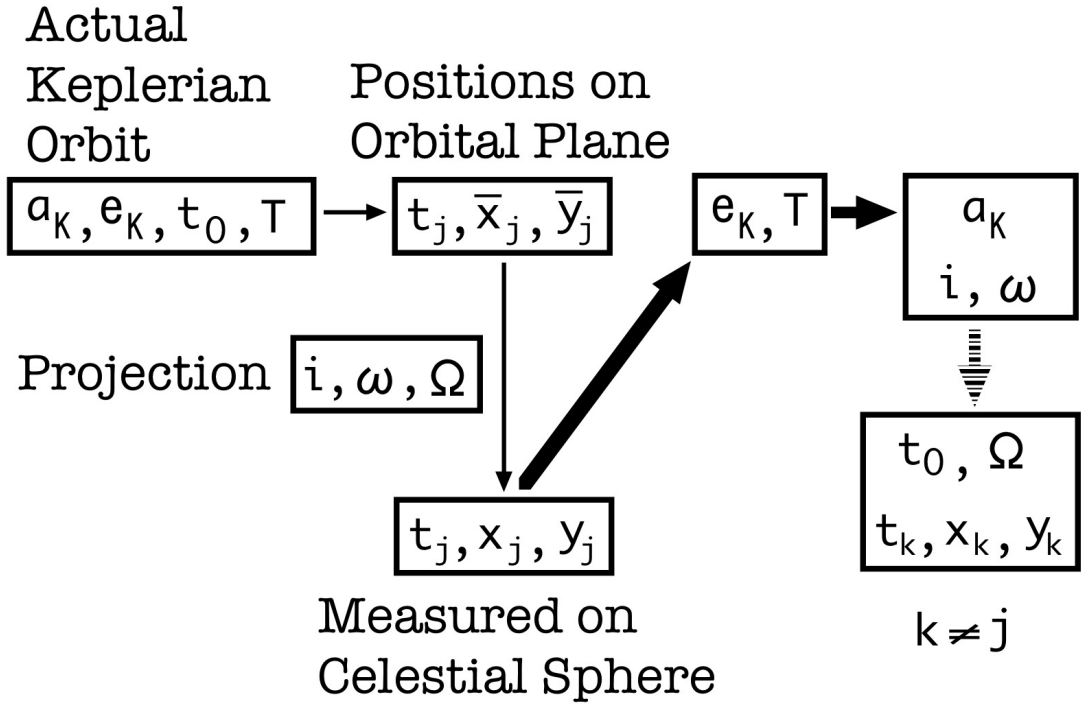}
  \end{center}
  \caption{Flow chart of our procedure of orbit determination. 
The thin arrow denotes purely theoretical steps, where we initially 
assume an actual Keplerian orbit parameterized by $(a_K, e_K, t_0, T)$. 
A star's position on the orbital plane at each time $t_j$ 
is projected onto the celestial sphere defined by $i$, $\omega$ and $\Omega$. 
The thick arrow denotes observational steps, where we start from measuring 
star's positions on the celestial sphere as $(t_j, {\bf x}_j)$. 
The steps of determining $e_K$, $a_K$, $T$ and $(i, \omega)$ have 
been examined (Asada et al. 2004, Asada 2007). 
The remaining steps of computing $t_0$, $\Omega$ and $(t_k, {\bf x}_k)$ 
in the simplified solution, denoted by the dashed arrow, 
are discussed in this paper.}
\label{fig2}
\end{figure}

\begin{figure}
  \begin{center}
    \FigureFile(130mm,130mm){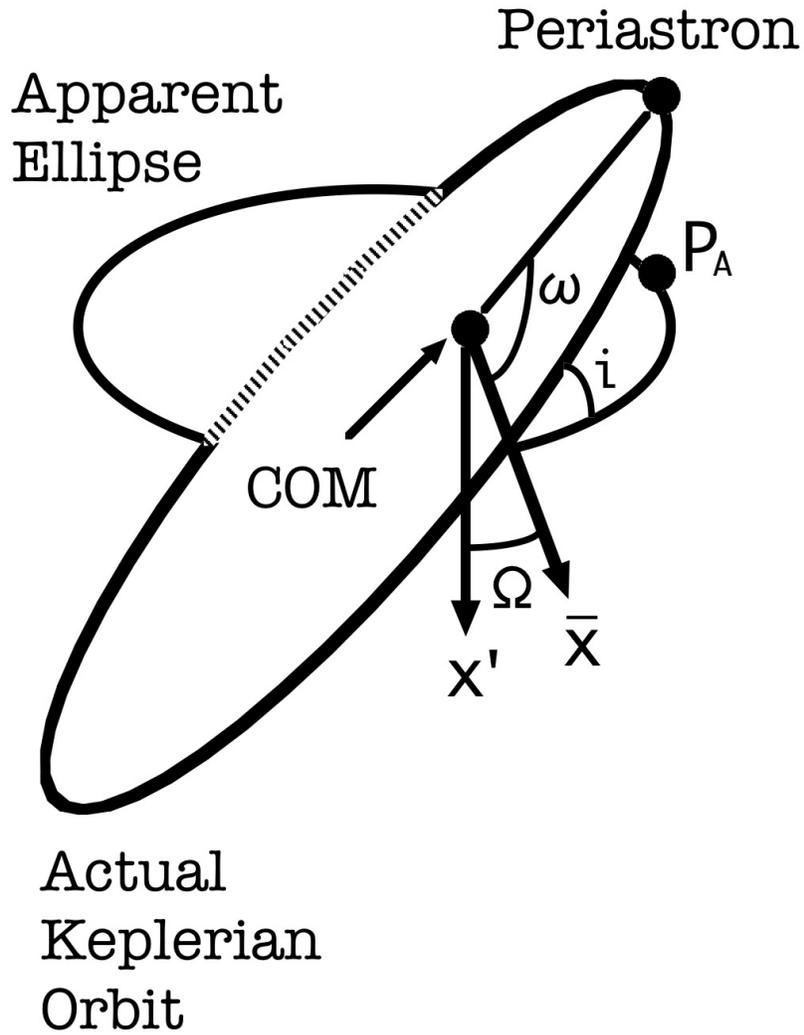}
  \end{center}
  \caption{Actual Keplerian orbit and apparent ellipse 
in three-dimensional space. 
We introduce the inclination angle $i$, 
the argument of periastron $\omega$ and the longitude of 
ascending node $\Omega$. 
These angles relate two coordinates $(x^{\prime}, y^{\prime})$ 
and $(\bar{x}, \bar{y})$, both of which choose the origin 
as the common center of mass. 
Here the $x^{\prime}$ axis is taken to lie along the major axis 
of the apparent ellipse, 
while the $\bar{x}$-axis is taken to lie along 
the direction of the ascending node.}
\label{fig3}
\end{figure}

\begin{figure}
  \begin{center}
    \FigureFile(160mm,1600mm){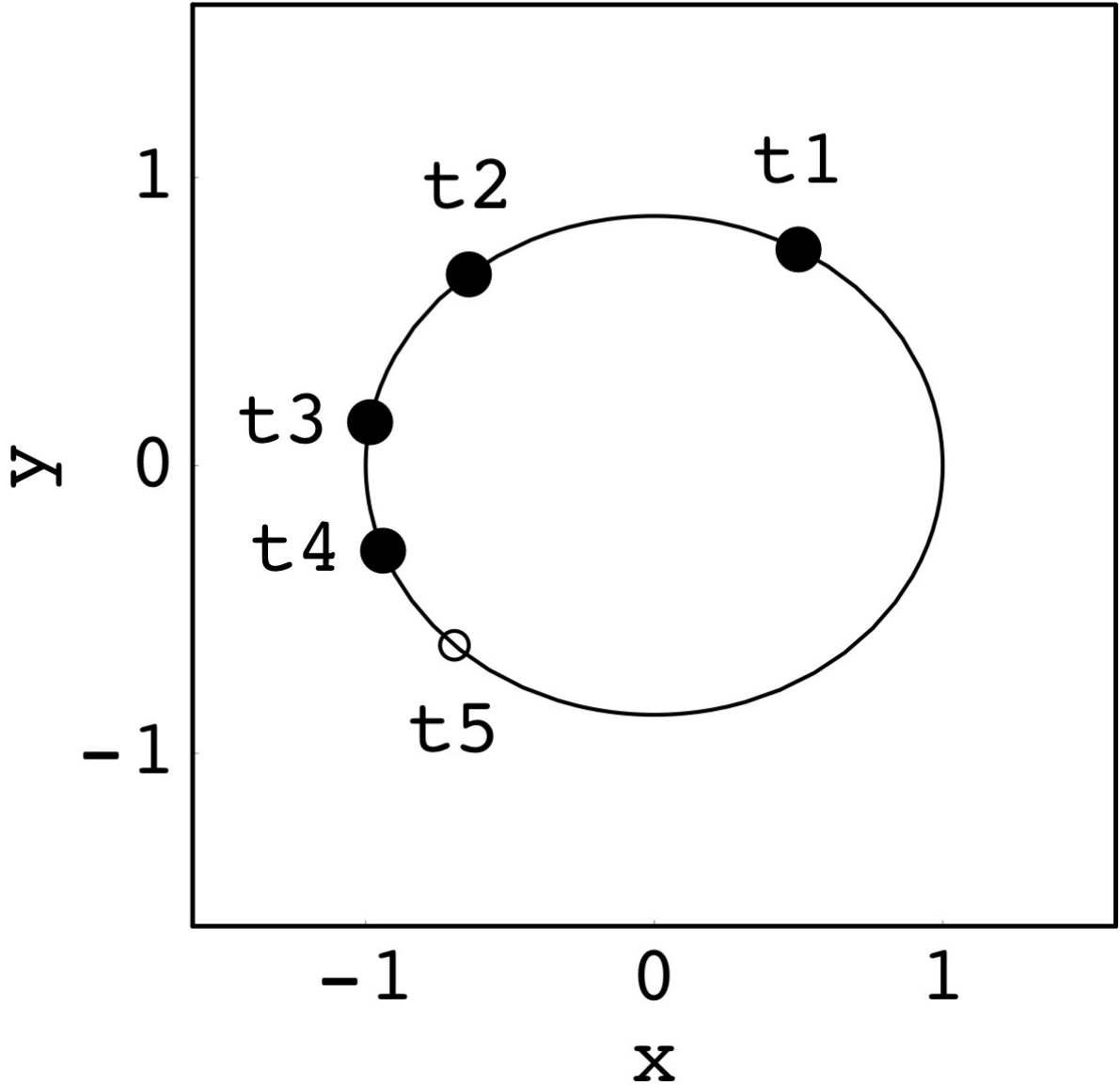}
  \end{center}
  \caption{Example of orbit determination. 
Here we assume $a=1$, $e=0.5$ as an apparent ellipse 
on the celestial sphere.  
Let a star located at $u_1=60$, $u_2=130$, $u_3=170$, $u_4=200$ 
(deg.) at each time $t_i$ for $i=1, 2, 3, 4$. 
Regarding time, we assume that 
the time interval $T(i+1, i)$ is the same as unity, 
namely $t_i = i-1$ (i.e. $t_1=0$), for simplicity. 
The simplified solution allows for arbitrary time interval.  
The observed positions of the star are denoted by the filled circle. 
The quantities determined in the present procedure are 
$e_K = 0.56$, $a_K = 1.2$, $i = 42$ deg. and $\omega = 79$ deg. 
We obtain $t_0 = 0.04$ by Eq. ($\ref{t0}$). 
The star's position at $t_5 = 5$ is obtained as $u_5=2.3 \times 10^2$ deg. 
The location is denoted by the circle. 
Using the determined $e_K$, $a_K$, $i$ and $\omega$, 
we determine $u_5$ at $t_5$ also by employing the conventional procedure
(indicated by the thin arrow in Fig. $\ref{fig2}$). 
In the latter case, we need to take account of the longitude of 
ascending node, $\Omega$. 
We obtain $\Omega = 15$ (deg.) by Eq. ($\ref{tanOmega}$). 
The results of $u_5$ by both methods agree with each other.}
\label{fig4}
\end{figure}

In this paper, we obtain in the simplified solution 
both the time of periastron passage
and the longitude of ascending node 
in order to complete the solution 
(See Fig. $\ref{fig2}$). 
In conclusion, the simplified solution requires neither iterative nor 
numerical methods when we determine all the elements 
including $t_0$ and $\Omega$. 
It does only when we wish to determine the star's position 
at arbitrary time. 

Before closing this paper, it is worthwhile to mention that 
Eqs. ($\ref{t0}$), ($\ref{tanOmega}$) and ($\ref{position}$) 
can be applied to a case of open orbits in the straightforward manner. 
For open orbits, expressions of $x_e$, $y_e$, $e_K$, $a_K$, $i$ 
and $\omega$ have been already derived
in the framework of the simplified solution (Asada 2007). 


We would like to thank Hiroshi Kinoshita for stimulating
conversations.
We wish to thank Toshio Fukushima 
for his useful comments on the earlier version of 
the manuscript. 
This work was supported by a Japanese Grant-in-Aid 
for Scientific Research from the Ministry of Education, 
No. 19035002.


\clearpage

\clearpage

\clearpage

\clearpage

\clearpage

\begin{table}
\caption{Numerical example of orbital elements. 
The reference direction is 
taken along the $x$-axis, which is different from the major axis 
of the apparent ellipse in this example. 
$\Omega$ is the angle measured from the reference direction. 
}
\label{table1}
  \begin{center}
    \begin{tabular}{lllllll}
$a_k$ & $e_k$ & $i$ & $\omega$ & $\Omega$ & $t_0$ & $T$ \\
\hline
1 & 0.5 & $\pi/8$ & $\pi/9$ & $\pi/10$ & 0 & 20 
    \end{tabular}
  \end{center}
\end{table}

\begin{table}
\caption{Virtual observational data set 
based on the orbital elements in Table 1. 
For simplicity, we assume that observations are sampled 
with the same frequency.} 
\label{table1}
  \begin{center}
    \begin{tabular}{lll}
$t_j = j$ & $x_j$ & $y_j$ \\
\hline
1 & 0.372003 & 0.838658  \\
2 & -0.0648698 & 0.831542  \\
3 & -0.404177 & 0.696231  \\
4 & -0.646827 & 0.509083  \\
5 & -0.809209 & 0.304280  \\
    \end{tabular}
  \end{center}
\end{table}

\end{document}